\documentclass[letterpaper, 10 pt, conference]{ieeeconf}  

\IEEEoverridecommandlockouts                              
\usepackage{amsmath} 
\usepackage{amssymb}  

\usepackage{url}
\usepackage[ruled, vlined, linesnumbered]{algorithm2e}
\usepackage{verbatim} 
\usepackage{soul, color}
\usepackage{lmodern}
\usepackage{fancyhdr}
\usepackage[utf8]{inputenc}
\usepackage{fourier} 
\usepackage{array}
\usepackage{makecell}
\usepackage{scicite}
\usepackage{float}
\usepackage{upgreek}
\usepackage{graphicx}
\usepackage{subfigure}
\usepackage{gensymb}

\UseRawInputEncoding

\SetNlSty{large}{}{:}

\makeatletter

\newcommand{\Rom}[1]{\expandafter\@slowromancap\romannumeral #1@}
\makeatother

\title{\LARGE
YIG/CoFeB bilayer magnonic diode}

\author{Noura Zenbaa,$^{1,2\ast}$ Khrystyna O. Levchenko,$^{1}$ Jaganandha Panda,$^{3}$ Krist\'yna Dav\'idkov\'a,$^{1,2}$\\ Moritz Ruhwedel,$^{4,5}$ Sebastian Knauer,$^{1}$ Morris Lindner,$^{6}$ Carsten Dubs,$^{6}$ Qi Wang,$^{7}$ Michal Urb\'anek,$^{3}$ \\ Philipp Pirro,$^{8}$ Andrii V. Chumak,$^{1,3\ast}$\\
\\
\normalsize{$^{1}$University of Vienna, Faculty of Physics, Vienna 1090, Austria}\\
\normalsize{$^{2}$University of Vienna, Vienna Doctoral School in Physics, Vienna 1090, Austria}\\
\normalsize{$^{3}$CEITEC BUT, Brno University of Technology, 612 00 Brno, Czech Republic}\\
\normalsize{$^{4}$Deutsches Zentrum f\"ur Luft- und Raumfahrt e. V. (German Aerospace Center, DLR),}\\
\normalsize{Institute of Solar Research
Linder H\"ohe, 51147 K\"oln, Germany}\\
\normalsize{$^{5}$Chair of Solar Technology
RWTH Aachen University
Linder H\"ohe, 51147 K\"oln, Germany}\\
\normalsize{$^{6}$INNOVENT e.V. Technologieentwicklung, 07745 Jena, Germany}\\
\normalsize{$^{7}$School of Physics, Hubei Key Laboratory of Gravitation and Quantum Physics,}\\ \normalsize{Institute for Quantum Science and Engineering, Huazhong University of Science and Technology, Wuhan, 430074, China.}\\
\normalsize{$^{7}$Fachbereich Physik and Landesforschungszentrum OPTIMAS, RPTU Kaiserslautern-Landau, 67663 Kaiserslautern, Germany}
\\
\normalsize{$^\ast$To whom correspondence should be addressed; E-mail: noura.zenbaa@univie.ac.at, andrii.chumak@univie.ac.at}}

\date{}

\begin{document}

\maketitle

\thispagestyle{plain}
\pagestyle{plain}

\begin{abstract}

We demonstrate a magnonic diode based on a bilayer structure of Yttrium Iron Garnet (YIG) and Cobalt Iron Boron (CoFeB). The bilayer exhibits pronounced non-reciprocal spin-wave propagation, enabled by dipolar coupling and the magnetic properties of the two layers. The YIG layer provides low damping and efficient spin-wave propagation, while the CoFeB layer introduces strong magnetic anisotropy, critical for achieving diode functionality. Experimental results, supported by numerical simulations, show unidirectional propagation of Magnetostatic Surface Spin Waves (MSSW), significantly suppressing backscattered waves. This behavior was confirmed through wavevector-resolved and micro-focused Brillouin Light Scattering measurements and is supported by numerical simulations. The proposed YIG/SiO$_2$/CoFeB bilayer magnonic diode demonstrates the feasibility of leveraging non-reciprocal spin-wave dynamics for functional magnonic devices, paving the way for energy-efficient, wave-based signal processing technologies.\\

\end{abstract}


\section{INTRODUCTION}\label{sec1}

Magnon-based devices have been of great interest in recent decades due to their unique ability to leverage magnons, the quanta of spin waves, as information carriers. One of the key advantages of magnons lies in their low-energy consumption, their frequencies that can reach the terahertz range~\cite{wu_high-performance_2017}, and their wavelengths that can reach down to the lattice constant~\cite{barman_2021_2021, chumak_advances_2022}. These properties make magnonic devices highly promising for achieving extreme miniaturization in future technologies. Furthermore, the tunability of spin-wave dispersion provides a powerful tool for engineering non-reciprocal behaviour, allowing spin waves to propagate with different characteristics in opposite directions. The non-reciprocal propagation of spin waves can be induced and controlled through various mechanisms, such as utilizing curvilinear 3D structures~\cite{hertel_curvature-induced_2013, pylypovskyi_coupling_2015, sheka_torsion-induced_2015, pylypovskyi_rashba_2016, otalora_curvature-induced_2016} or employing magnetic systems dominated by the Dzyaloshinskii-Moriya interaction (DMI)~\cite{dzyaloshinsky_thermodynamic_1958, moriya_anisotropic_1960, 
nembach_linear_2015, tacchi_interfacial_2017, santos_nonreciprocity_2020}. Another method is the use of magnetic bilayers in a parallel magnetization state~\cite{gerevenkov_nonreciprocal_2023, gallardo_spin-wave_2019}, antiparallel magnetization state~\cite{grunberg_layered_1986, gallardo_reconfigurable_2019} or in the form of a synthetic antiferromagnet~\cite{verba_wide-band_2019, wojewoda_unidirectional_2024}. For example, a CoFeB/Py exchange-coupled bilayer was demonstrated numerically and experimentally as a magnonic diode, where the exchange coupling facilitated non-reciprocal propagation~\cite{grassi_slow-wave-based_2020}. Spin-wave non-reciprocity is particularly beneficial for developing functionalities such as magnonic isolators, circulators, and diodes, which are essential components in advanced signal processing and computing systems~\cite{verba_conditions_2013, yu_room-temperature_2016, gladii_frequency_2016}. These attributes position magnon-based devices as a compelling platform for energy-efficient, scalable, and reconfigurable information technologies.

Here, we present the experimental realization of a magnonic diode based on a magnetic bilayer of Yttrium Iron Garnet/Cobalt Iron Boron (YIG/CoFeB), coupled via dipolar interaction through a non-magnetic spacer of SiO$_2$. Unlike exchange-coupled systems such as CoFeB/Py bilayers~\cite{grassi_slow-wave-based_2020}, the dipolar coupling in our system preserves the low damping of YIG, with values remaining in the range of 10$^{-4}$, facilitating extended spin-wave propagation lengths. The broken spatial symmetry along the thickness causes non-reciprocal propagation when Magnetostatic Surface Spin Waves (MSSWs) are excited by transverse in-plane magnetization. The spin-wave non-reciprocity arises from the dynamic dipolar field interaction between the two magnetically heterogeneous layers, which exhibit distinct magnetic properties such as saturation magnetization $M_s$, propagation frequency $f$, phase $\phi$ and group velocities. Under these conditions, the spin waves propagating in the two layers, with different frequencies and phases, create stray fields that interact dynamically. These stray fields interact with the dynamic magnetization in each layer. This interaction modifies the dynamic dipolar interaction energy density, 
$\epsilon_d = -\frac{\mu_0}{2} \mathbf{m} \cdot \mathbf{h}_{\mathrm{stray}}$~\cite{gallardo_reconfigurable_2019}, making it non-reciprocal with respect to the spin-wave propagation direction. The energy density is minimized when $m$ and $h^{\mathrm{stray}}$ are parallel and maximized when they're antiparallel. Larger disparities between the magnetic properties of the two layers amplify this non-reciprocity, which was validated with numerical simulations and experimentally using wavevector-resolved Brillouin Light Scattering ($k$-BLS) spectroscopy for non-reciprocal dispersion investigation and micro-focused $\upmu$-BLS for spin-wave transport measurements.

This pronounced non-reciprocal behavior underscores the potential of YIG/CoFeB-based bilayers for magnonic isolators, phase shifters, and other wave-based signal processing devices.

\section{Methods and Materials}\label{sec2}
\subsection{Materials}\label{subsec2}
The YIG film used is 100\,nm-thick grown on a 500\,$\upmu$m-thick Gadolinium Gallium Garnet (GGG) substrate using Liquid Phase Epitaxy (LPE)~\cite{dubs_sub-micrometer_2017, dubs_low_2020}. We used magnetron sputtering to grow a 5\,nm-thick SiO$_2$ (nonmagnetic spacer ) and the ferromagnetic layer of 40\,nm-thick CoFeB on the YIG film - see Fig.~\ref{fig1}(a). This magnetic stack of YIG(100\,nm)/SiO$_2$(5\,nm)/CoFeB(40\,nm) was characterized using Vector Network Analyzer Ferromagnetic Resonance (VNA-FMR) spectroscopy~\cite{maksymov_broadband_2015} to extract the YIG material parameters within the stack which showed an effective magnetization $M_{\mathrm{eff}} = (160.1 \pm 0.3$)\,mT, inhomogeneous linewidth broadening $\mu_0 \Delta H_0 = (0.161 \pm 0.02)$\,mT and Gilbert damping coefficient $\alpha = (4.4 \pm 0.02) \times 10^{-4}$. A plain YIG film cut from the same sample wafer was characterized using VNA-FMR spectroscopy and a Gilbert damping coefficient of $\alpha = (7.7 \pm 0.65) \times 10^{-5}$ was measured, which is almost one order of magnitude smaller than that in the magnetic bilayer. The damping in the YIG/CoFeB bilayer (without a nonmagnetic spacer) was investigated, showing a significant broadening of the linewidth: 1.12\,mT at an external field of 10\,mT compared to 0.18\,mT at 14\,mT in the YIG/SiO$_2$/CoFeB bilayer. This behavior qualitatively aligns with the findings in~\cite{ qin_exchange-torque-induced_2018}, where direct exchange coupling between YIG and a ferromagnetic metallic layer facilitates dynamic exchange torque at the interface. The enhanced damping in such bilayers arises from this exchange interaction, which not only transfers angular momentum but also allows for the efficient excitation of higher-order standing modes, particularly near the ferromagnetic resonance frequencies of either layer constituting the bilayer.

To excite and detect coherent spin waves using $\upmu$-BLS, we fabricated a 600-nm-wide microwave strip antenna via electron-beam lithography. The antenna is composed of a Ti(5\,nm)/SiO$_2$(10\,nm)/Ti(5\,nm)/Au(80\,nm) stack deposited by electron beam evaporation. 

\subsection{Micromagnetic simulations}

To choose the material and thickness of the second magnetic layer of the bilayer, a micromagnetic simulation study was performed using the GPU-accelerated program MuMax$^3$~\cite{vansteenkiste_design_2014}. For the nanometer-thick bilayer of YIG/SiO$_2$/CoFeB, a length $x = 40\,\upmu$m, width $y =$ 100\,nm, and thickness $t = $145\,nm are considered. The periodic boundary conditions PBC = (10, 100, 0) are applied along the length and width to mimic a thin plane film. The cell size was set to (10\,nm$ \times $10\,nm) in the in-plane direction and to 5\,nm in the out-of-plane direction, to account for the nonmagnetic spacer SiO$_2$. The material parameters used for YIG are saturation magnetization $M_s = 140$\,kA/m (without taking into account the anisotropy fields), exchange constant $A_{\mathrm{ex}} =3.5$ \,pJ/m and $\alpha = 2\times10^{-4}$, for SiO$_2$ are $M_s = 0$\,kA/m,  $A_{\mathrm{ex}} = 0$ \,pJ/m and $\alpha = 1$ and lastly for CoFeB are $M_s = 1250$\,kA/m, $A_{\mathrm{ex}} =15$ \,pJ/m and $\alpha = 4\times10^{-3}$. A sinc-function pulse in time excites spin waves in the in-plane transverse direction.

\subsection{BLS measurements}

BLS spectroscopy~\cite{sebastian_micro-focused_2015} relies on the process of inelastic scattering of light as a result of photon-magnon interaction, while conserving the in-plane momentum. This inelastic light-scattering can create or annihilate a magnon, resulting in Stokes and anti-Stokes peaks in the BLS spectrum. The momentum of detected magnons is given by $k_{\mathrm{magnon}} = \frac{4\pi}{\lambda_L} \sin\theta$, where $\lambda_L$ is the incident laser wavelength and $\theta$ is the incident angle of the laser. The frequencies of the inelastically scattered light are analyzed using a Tandem Fabry-P\'erot Interferometer (TFPI)~\cite{mock_construction_1987}.

To validate the micromagnetic simulations and the non-reciprocal behavior of the magnetic bilayer, $k$-resolved BLS measurements were performed. The measurements are done in the Damon-Eshbach configuration to measure MSSWs where the bias magnetic field is applied along the sample plane but perpendicular to the incident laser beam plane. The monochromatic laser used is of wavelength $\lambda_L = 491$\,nm. The $k$ resolution comes from changing the incidence angle of the laser beam in the range of 2.5$\degree$ to 80$\degree$ which corresponds to $k_{\mathrm{magnon}} =$ 1.1163\,rad/$\upmu$m to 25.205\,rad/$\upmu$m. The thermal signal was recorded for both polarities of the bias field of 200\,mT.

The micro-focused $\upmu$-BLS is used to measure coherent spin waves and validate non-reciprocal propagation. A continuous-wave,
monochromatic polarized laser beam of $\lambda_L = 457$\,nm is used. The laser is focused through the GGG substrate onto the YIG film, using a cover glass corrected objective.  This objective lens, with a numerical aperture (NA) of 0.85, achieves a focal spot diameter $d$ of 330\,nm when used with $\lambda_L$. This configuration enables the detection of wavevectors up to 12\,rad/$\upmu$m~\cite{wojewoda_modeling_2024}.

\section{RESULTS AND DISCUSSION}

\begin{figure}
\centerline{\includegraphics[width=18.5pc]{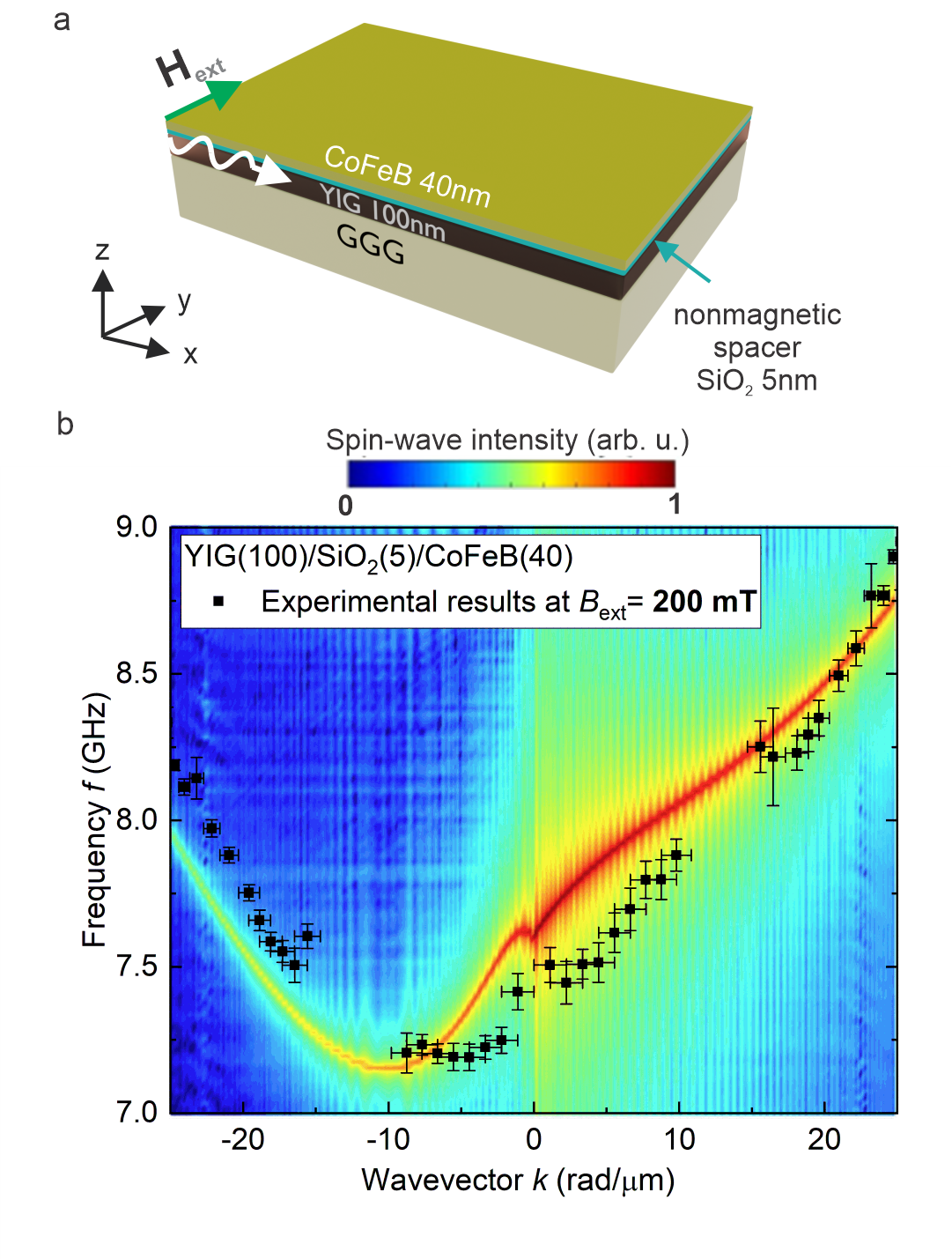}}
\caption{(a) Schematics of the nano-thick magnetic bilayer of GGG/YIG(100\,nm)/SiO$_2$(5\,nm)/CoFeB(40\,nm). (b) Dispersion relations of Magnetostatic Surface Spin Waves (MSSWs) in the magnetic bilayer film simulated numerically using MuMax$^3$, shown as a color map, where red is maximum spin-wave intensity and blue is zero intensity. The square symbols represent the experimental data of spin-wave frequency $f$ as a function of wavevector $k$ measured on the same bilayer film with wavevector-resolved Brillouin Light Scattering ($k$-resolved BLS), including the error bars.}
\label{fig1}
\end{figure}
The numerical simulation of the nanometer-thick magnetic bilayer GGG/YIG(100\,nm)/SiO$_2$(5\,nm)/CoFeB(40\,nm) is shown in Fig.~\ref{fig1}(b) as a color map. It depicts the dispersion relations of MSSWs in the YIG film as part of the stack at a 200\,mT bias field. The dispersion curve shows pronounced frequency non-reciprocity for counterpropagating surface waves. Specifically, in the $-k$ propagation direction, the surface waves behave like Backward-Volume Magnetostatic Spin Waves (BVMSWs), where the slope of the dispersion curve is negative in the small $k$ limit and becomes zero around $k = -10$\,rad/$\upmu$m. 
In this range, the group velocity is negative and reaches zero at $k = -10$\,rad/$\upmu$m, while the phase velocity remains positive. This results in the group velocity and phase velocity pointing in opposite directions for wavevectors between $-10$\,rad/$\upmu$m and $0$\,rad/$\upmu$m. Such behavior highlights the unique non-reciprocal spin-wave dynamics in the bilayer structure. These simulations suggest that in this magnetic bilayer, MSSWs of wavevectors $k = \pm10$\,rad/$\upmu$m are unidirectional. By adjusting the thickness of the CoFeB magnetic layer, we were able to achieve such a flat frequency plateau in only one direction~\cite{grassi_slow-wave-based_2020}. Similar to the phenomenon measured in Synthetic Antiferromagnets (SAFs)~\cite{wojewoda_unidirectional_2024}, we deduce from our simulations that zero momentum waves in the bilayer can propagate as they don't exhibit zero group velocities.

\begin{figure}
\centerline{\includegraphics[width=18.5pc]{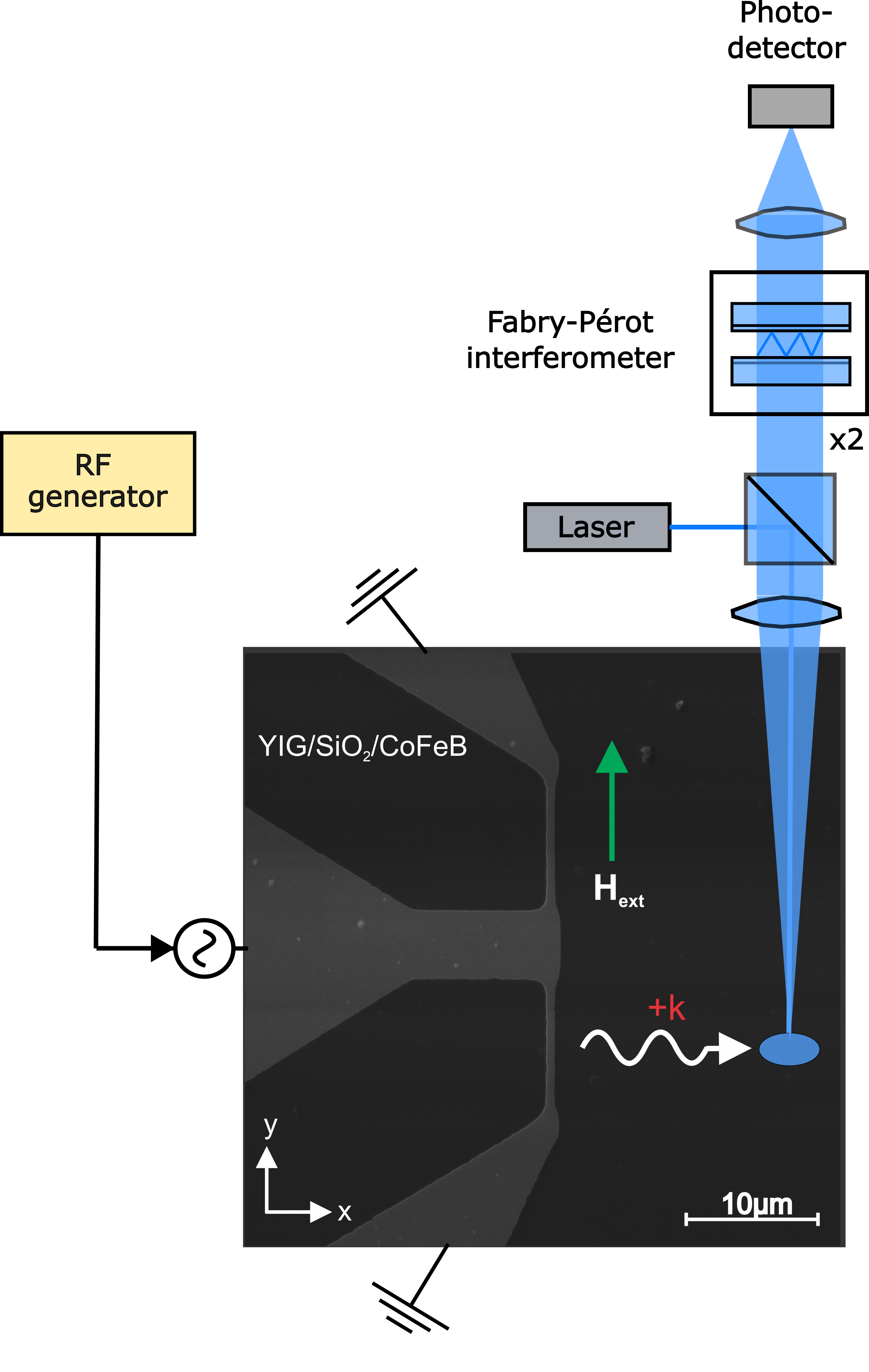}}
\caption{Schematics of the experimental setup micro-focused ($\upmu$-BLS) and scanning electron microscopy (SEM) micrograph of the structure under investigation.A magnetic bias field of $\mu_0 \mathrm{H_{ext}} =$ 200\,mT is applied perpendicular (along the y-direction) to the wave propagation (along the x-direction). Spin waves are excited through a continuous RF signal of power $P = $ 5\,dBm into the microstrip antenna.}
\label{fig2}
\end{figure}

\begin{figure*}
\centerline{\includegraphics[width=43pc]{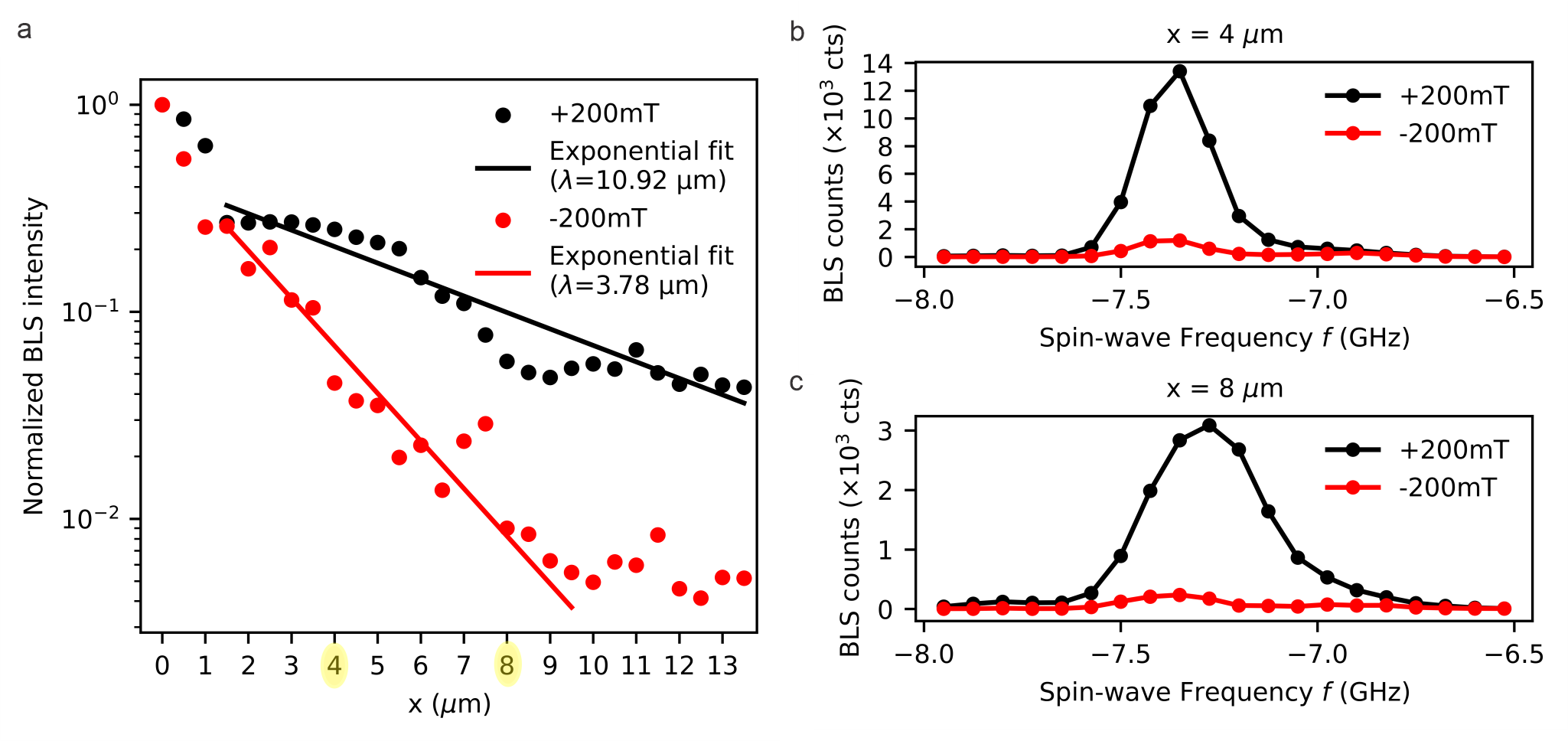}}
\caption{(a) Normalized BLS intensity as a function of laser position in the x-direction with regards to the antenna position dependent on the applied bias field for an excitation frequency of 7.23\,GHz. Spin wave spectra measured at distances of (b) 4\,$\upmu$m and (c) 8\,$\upmu$m from the microstrip antenna.}
\label{fig3}
\end{figure*}

The dispersion curve of MSSW in the fabricated magnetic bilayer film was measured using $k$-resolved BLS. The experimental results are shown as the square symbols in Fig.~\ref{fig1}(b) on top of the simulated color map. There is an agreement between the experimental data and the simulations with a slight shift in frequencies that can be explained by the fact that the fabricated bilayer thicknesses could be slightly inaccurate. Nevertheless, the non-reciprocal dispersion behavior is conserved and matches very well what we found in simulations. The gaps seen in the experimental data correspond to the phonon modes measured in the sample. As mentioned above, the data was collected by measuring the spin-wave signal while varying the incident laser angle with respect to the sample plane at both polarities of the magnetic field ($\pm$200\,mT). While it is typically sufficient to measure the Stokes and anti-Stokes peaks at one field polarity to capture the spin waves propagating in opposite directions, measurements at both polarities were performed here because the reference signal position shifted slightly when the field polarity was reversed. By combining data from both polarities, this reference shift could be compensated, allowing for more accurate error estimation.

The schematic of the $\upmu$-BLS experimental setup together with an SEM image of the sample is shown in Fig.~\ref{fig2}. We used the microstrip microwave antenna on the nanometer-thick magnetic bilayer film to excite coherent spin waves propagating perpendicular to the applied bias magnetic field of $\pm$200\,mT. We applied a continuous wave RF signal of power $P =$ 5\,dBm through the microstrip antenna. The setup uses a monochromatic blue laser that is focused on the sample through an objective lens. A polarized beam cube is used to pass only the photons which were inelastically scattered on magnons to the six-pass TFPI which uses two mirror pairs. After the frequencies are analyzed in the TFPI, the light is guided to a photo-detector.

We performed the measurements using the $\upmu$-BLS in the form of a 1D line scan along the x-direction that starts at the antenna position with 500\,nm steps. The 1D line scan was repeated for the two polarities of the magnetic field $\pm$200\,mT on the same side of the antenna shown in Fig.~\ref{fig2}. The results of the line scans are shown in Fig.~\ref{fig3}(a) in terms of normalized BLS intensity as a function of x position with regard to the antenna position at an excitation frequency of 7.23\,GHz. This excitation frequency was selected as it generated the strongest coherent signal observed during the measurements. The BLS intensity is proportional to the spin-wave intensity. We normalized each line scan to its own to eliminate the non-reciprocity effect caused by the non-reciprocal excitation efficiency characteristic of the Damon-Eshbach configuration.The position x $= 0$\,$\upmu$m is at the bottom of the antenna and x $= 0.5$\,$\upmu$m is just outside the 600\,nm-wide antenna. The decay length $\lambda$ was extracted by fitting the data to an exponential decay model $I = I_0 \exp(-2x/\lambda)$. The first three data points were excluded from the fit of both polarity line scans as they correspond to positions directly under or very near the antenna. These points are influenced by both propagation directions and potentially non-resonant modes, which contribute to the measured counts. We excluded the data points after x =$10 \upmu$m of the $-200$\,mT field line scan as well since they reach the thermal level. The decay length $\lambda$ of spin waves in the positive field direction was measured to be nearly three times that of spin waves in the opposite field polarity, with values of $\lambda_{\mathrm{+200\,mT}} = 10.92 \upmu$m and $\lambda_{\mathrm{-200\,mT}} = 3.78 \upmu$m. 

Fig.~\ref{fig3}(b) and \ref{fig3}(c) show the spin-wave spectra measured at 4$\upmu$m and 8$\upmu$m away from the antenna, respectively. The spin waves excited at a $-200$\,mT bias field drop to below 1000 counts compared to almost 14$\times$10$^3$ at the opposite polarity field - see the spin-wave spectrum in Fig.~\ref{fig3}(b). Further away from the antenna at x $= 8$\,$\upmu$m, the spin waves are at the thermal level with counts way below 500 at $-200$\,mT while at $+200$\,mT, spin waves were still detected with more than 3000 counts as shown in Fig.~\ref{fig3}(c). These results demonstrate the unidirectional nature of spin-wave propagation in our nanometer-thick magnetic bilayer, with significantly longer decay lengths at  $+200$\,mT compared to  $-200$\,mT. The bilayer's ability to support long-distance spin-wave propagation (over $10 \upmu$m) and its strong asymmetry in propagation directions establish it as an effective magnonic diode, capable of directional spin-wave transport.

\section{Conclusion}\label{sec5}
This study demonstrates a YIG/SiO$_2$/CoFeB bilayer-based magnonic diode, leveraging non-reciprocal spin-wave dynamics driven by dipolar interactions and the distinct magnetic properties of the layers. Through wavevector-resolved and micro-focused Brillouin Light Scattering measurements, we confirmed the unidirectional propagation of Magnetostatic Surface Spin Waves and the suppression of backward waves. Spin waves in the $+200$\,mT field direction exhibited a decay length exceeding $10 \, \upmu\mathrm{m}$, nearly three times longer than the decay length measured at $-200$\,mT ($\lambda_{\mathrm{-200\,mT}} = 3.78 \, \upmu\mathrm{m}$), highlighting the bilayer's efficiency in supporting long-distance directional spin-wave transport. Numerical simulations further supported these findings, emphasizing the bilayer's potential for energy-efficient, wave-based signal processing applications, including magnonic diodes, isolators, and phase shifters. These results provide a foundation for the development of scalable magnonic devices for next-generation computing technologies.

\section*{Acknowledgements}

The financial support by the Austrian Science Fund (FWF) MagFunc [10.55776/I4917] is acknowledged. A.C. acknowledges the financial support by the European Research Council (ERC) Proof of Concept Grant 101082020 5G-Spin. S.K. acknowledges the support by the H2020-MSCA-IF under Grant No. 101025758 ("OMNI"). The work of M.L. was supported by the German Bundesministerium für Wirtschaft und Energie (BMWi) under Grant No. 49MF180119. M.U. acknowledges the support from project No. CZ.02.01.01/00/22008/0004594 (TERAFIT). CzechNanoLab project LM2023051 is acknowledged for the financial support of the sample fabrication at CEITEC Nano Research Infrastructure.

\bibliography{ref}

\bibliographystyle{Science}

\end{document}